# Soft Porous Crystals:
# Extraordinary Responses to Stimulation

Chimie ParisTech, PSL University, CNRS, Institut de Recherche de Chimie
François-Xavier Coudert*

Received April 1, 2018; E-mail: fx.coudert@chimieparistech.psl.eu

Recent years have seen a large increase in the number of reported framework materials, including the nowadays-ubiquitous metal–organic frameworks (MOFs). Many of these materials show flexibility and stimuli-responsiveness, i.e. their structure can undergo changes of large amplitude in response to physical or chemical stimulation. We describe here a toolbox of theoretical approaches, developed in our group and others, to shed light into these materials' properties. We focus on their behavior under mechanical constraints, temperature changes, adsorption of guest molecules, and exposure to light. By means of molecular simulation at varying scale, we can now probe, rationalize and predict the behavior of stimuli-responsive materials, producing a coherent description of soft porous crystals from the unit cell scale all the way to the behavior of the whole crystal. In particular, we have studied the impact of defects in soft porous crystals, and developed a methodology for the study of their disordered phases (presence of correlated disorder, MOF glasses, and liquid MOFs).

■■ 1. Introduction

Recent years have seen a large increase in the number of reported framework materials, including the nowadays-ubiquitous metal–organic frameworks (MOFs), but also covalent organic frameworks, dense coordination polymers, and supramolecular frameworks. Many of these materials show flexibility and stimuli-responsiveness, i.e. their structure can undergo changes of large amplitude in response to physical or chemical stimulation.[1,2] Professor Kitagawa has defined these *"soft porous crystals"* (SPCs) as *"dynamic frameworks that are able to respond to external stimuli such as light, electric fields or the presence of particular species, […] and can change their channels reversibly while retaining high regularity"*.[3] Such systems are widely studied, not only for their challenge of fundamental understanding of their behavior, or the beauty of their structures, but also because their stimuli-responsiveness make them great targets for applications. Once an external constraint is applied, the structure of the soft porous crystal changes, and this in turn affects its physical and chemical properties. To give only one striking example, Lyndon et al.[4] reported the photoresponsive material Zn(AzDC)(4,4′-BPE)$_{0.5}$[5] where exposure to ultraviolet light can be used to trigger the uptake and release of carbon dioxide. SPCs thus display a change of their properties in response to their environment, making them *multifunctional materials*. It is thus expected that they can find applications as nanosensors, actuators, for targeted drug releases, and in other areas.

In this account, we describe here a toolbox of theoretical approaches, developed in our group and others throughout the world, to shed light into these materials' properties. For a background on the computational description of metal–organic frameworks, we refer the reader to the general reviews on the topic, such as Refs. 6 and 7. We focus here specifically on the theoretical description of the behavior of MOFs under mechanical constraints, temperature changes, adsorption of guest molecules, and exposure to light. By means of molecular simulation at varying scale, we can probe, rationalize and predict the behavior of stimuli-responsive materials, producing a coherent description of soft porous crystals from the unit cell scale all the way to the behavior of the whole crystal. In particular, we have studied the impact of defects in soft porous crystals, and developed a methodology for the study of their disordered phases (presence of correlated disorder, MOF glasses, and liquid MOFs).

■■ 2. Microscopic mechanisms of MOF flexibility

2.1. The link between adsorption and mechanics

Historically, the first examples of flexible MOFs where observed upon adsorption or desorption of guest molecules, whether they were solvent molecules, external gas or liquid. Among the most famous of these early examples, we can cite the MIL-53 family of materials,[8] MIL-88,[9] ELM-11,[10] Co(BDP),[11] Cu(4,4′-bipy)(dhbc)$_2$,[12] and DMOF-1.[13] In parallel with the development of

Corresponding Author: François-Xavier Coudert
Address: 11 rue Pierre et Marie Curie, 75005 Paris, France
https://www.coudert.name/

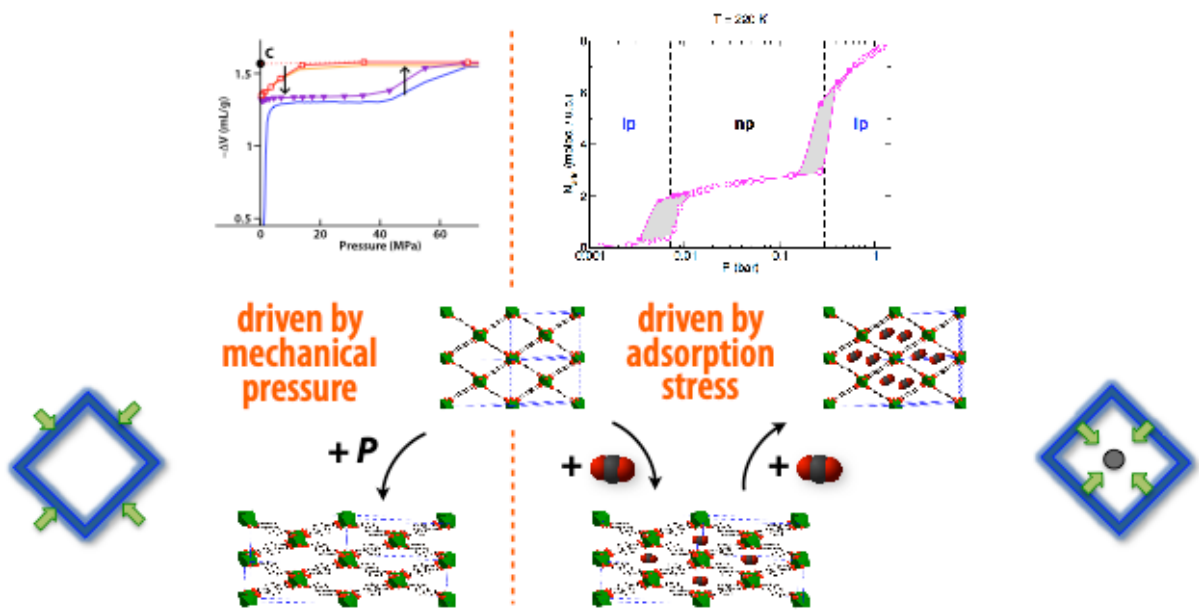

**Fig. 1** Representation of the "breathing" structural transition in the MIL-53 framework, which can be driven by adsorption (right) and mechanical compression (left). Insets on top: volume–pressure curve from mercury compression experiment (left), and Xe adsorption isotherm at 220 K (right). Reproduced with permission from Ref. 31. Copyright 2011 American Chemical Society.

experimental techniques for *in situ* characterization of this flexibility, we focused on the development of novel computational chemistry tools for modelling this flexibility, focusing in particular on the thermodynamics of the equilibrium between the phases. This lead to the development of free energy techniques (using Wang–Landau methods,[14,15] flat-histogram or transition matrix Monte Carlo,[16,17] or other free energy calculations[18,19]) to describe this competitive adsorption between the phases, and the emergence of complex phase $(P, T)$ diagrams for the materials,[20] where $P$ is the fluid pressure and $T$ the temperature — or $(P, T, x)$ for coadsorption of fluid mixtures, where $x$ is the composition.[21] This provided a successfully description of the driving forces towards the preferential adsorption in each possible phase of the material; in other words, a *thermodynamic* picture. It took into account the adsorption process, as well as temperature effects, by accounting for the energy and entropy of the framework, the adsorbate, and — most importantly — their coupling.

At the same time, experimental work reported in the literature started to demonstrate that this potential for flexibility of MOFs could also be triggered by a physical stimulus of a different nature: the application of mechanical pressure. This was demonstrated, for example, on MIL-53 by application of isostatic pressure by compression in liquid mercury, which triggered a reversible structural transition under compression.[22] There, the application of pressure is the driving force for the transition from an open, porous phase to a denser phase, with small unit cell and pore volume. Other authors later studied a variety of different flexible porous frameworks and pressure-transmitting fluids.[23,24,25,26] These transitions are part of the broader picture of MOF responses to pressure, which is very varied.[27]

However, MOF transitions induced by adsorption and mechanical constraints are inherently linked, as are their microscopic mechanisms. In fact, adsorption of guest molecules in microporous matrices creates a stress on the host framework, known as adsorption stress.[28,29,30] Unlike pressure, which is isotropic in a hydrostatic fluid, adsorption stress is anisotropic in nature, and its extent depends on the host framework, its loading, and the host–guest interactions. While this adsorption stress only creates small-scaled strain in common inorganic porous materials such as zeolites, in soft porous crystals the deformation can be of large amplitude. A detailed study on the archetypal MIL-53 "breathing" framework, for which a lot of experimental data is available, was performed by reinterpreting revisiting experimental data on mercury intrusion and *in situ* X-ray diffraction measurements during $CO_2$ adsorption.[31] We concluded there that the magnitude of the adsorption stress, exerted inside the pores by guest molecules, in order to induced the breathing transition, corresponds to the magnitude of external pressure applied from the outside the crystal in compression experiments. Thus, despite the difference in origin and nature of the stimulation, the microscopic mechanism is the same (see Figure 1). The structural transition occurs when the stress on the framework reaches a critical value, that the framework cannot resist anymore (the limit of mechanical stability).[32]

### 2.2. Mechanical properties of soft porous crystals

Because of the link described above between adsorption- and pressure-induced structural transitions in soft porous crystals, it is of high importance to know their mechanical properties in order to shed light into their behavior under stimulation. The past few years have seen a large research effort focused on the characterization of the mechanical properties of MOFs in general, and flexible frameworks in particular. Among the first works to quantify the "softness" of these materials were the works of Tan et al,[33] who studied the low shear modulus (among other elastic constants) of ZIF-8; and the studies by Bennett et al,[34,35,36] measuring the mechanical properties of zeolitic imidazolate frameworks (ZIFs) and their resistance to pressure-induced amorphization.

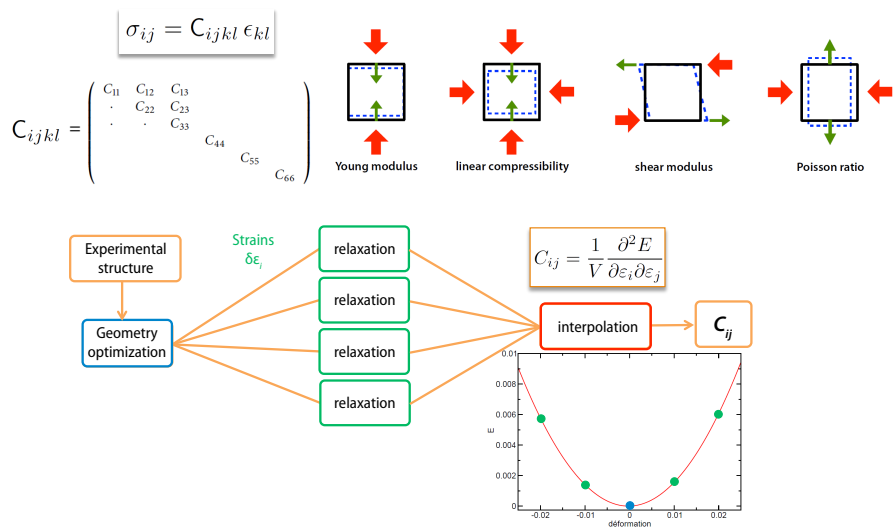

**Fig. 2** Determination of the second-order elastic constants $C_{ij}$ from a series of quantum chemistry calculations. Top right: graphical representation of the physical properties that can be derived from the stiffness tensor: Young's modulus, linear compressibility, shear modulus, and Poisson's ratio (red arrows represent applied stress, green arrows represent the strain measured in response).

In computational schemes, the most commonly used way to characterize the mechanical properties of a crystal is the determination of its stiffness tensor, or matrix of second-order elastic constants $C_{ij}$. These values can be calculated at the quantum chemistry level with high accuracy, by characterizing various strained structures, giving elastic constants in the "zero Kelvin" limit, i.e. without taking into account thermal motions. Analysis of these elastic constants can then be used to obtain more physically relevant elastic properties, including Young's modulus, shear modulus, linear compressibility and Poisson's ratio (described in Figure 2). We have, for example, used this scheme to characterize a series of SPCs, and demonstrate the existence of a "key signature" of the flexible nature of these frameworks, which can be seen in their elastic properties. All SPCs studied showed highly anisotropic elastic behavior (up to a 400:1 ratio) as well as the existence of some deformation modes exhibiting very low Young's modulus and shear modulus (over the order of ~0.1 GPa).[37,38,39] Other works using density functional theory (DFT) calculations of elastic constants have focused on properties such as auxeticity (negative Poisson's ratio), anisotropic elastic properties, etc.[40,41,42]

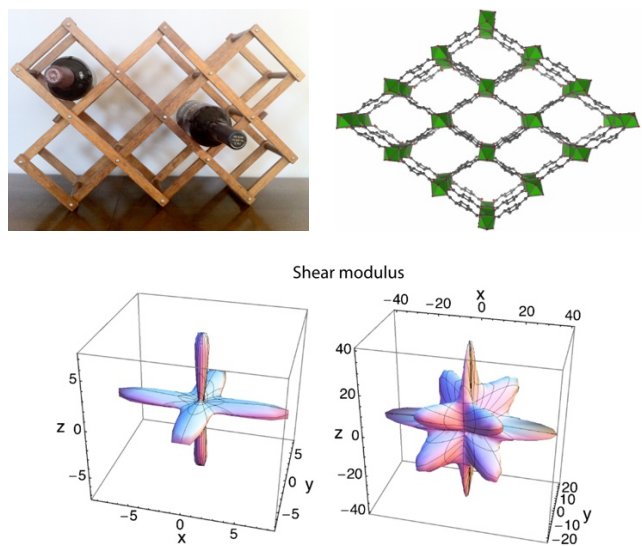

**Fig. 3** Top: "wine rack"-type framework of soft porous crystal MIL-53. Bottom: highly anisotropic shear modulus of MIL-53(Al), represented as 3D surfaces, in units of GPa. Reproduced with permission from Ref. 37. Copyright 2012 by The American Physical Society

Mechanical properties of porous frameworks can also be calculated at finite temperature with molecular dynamics, using either a stress–strain approach, or by averaging the fluctuations of the unit cell over a long periodic time. The computational cost of such schemes, however, means that they can only be performed by relying on "classical" simulations, where the interatomic interactions are described by an *ad hoc*, parameterized force field. The accuracy is thus lower, especially since elastic properties are second derivatives of the energy and thus quite sensitive. This approach has nevertheless been used in our work, for example in a study of the amorphization mechanism of ZIF-8 and ZIF-4, showing that pressure-induced amorphization in these materials is linked to a shear-mode softening of the material under pressure, which results in mechanical instability at moderate pressure (0.34 GPa).[43] We also relied on classical simulations of mechanical properties in a series of ZIFs of identical chemical composition, but varying topology, in order to demonstrate the influence of framework topological on the mechanical stability and thermal properties of soft porous crystals.[44]

### 2.3. Non-linear phenomena

We have described above the study of mechanical properties of soft porous crystals, from the perspective of their linear elastic regime. We showed that this linear behavior is, in many cases, tightly linked to the large-scale response observed under stimulation, as is observed in the case of the "wine rack" breathing MOFs. Even phenomena that are intrinsically nonlinear, such as the pressure-induced amorphization, can be studied to some extent as the limit of some linear elastic behavior. However, there is also an existing need for computational tools that can describe non-linear phenomena induced by pressure or adsorption. We present here briefly three such cases.

The first one is the case of the pressure response of zinc alkyl gate (ZAG) materials. Synthesized in the Clearfield group,[45] the ZAGs are zinc-based materials with alkylphosphonates as organic linkers; e.g., ZAG-4 has 1,4-butanebisphosphonate linkers. These materials, studied under high pressures (up to 10 GPa) using *in situ* single crystal X-ray diffraction, show nonlinear behavior with domains of both positive and negative linear compressibility along the *b* crystallographic axis. Although the structures feature a "wine rack" motif, we could show that their elastic behavior did not explain their unusual properties. In order to better understand the influence of pressure, it was necessary to perform enthalpy minimizations at increasing values of pressure, in order to track their structural evolution (in the zero Kelvin approximation).[46] This revealed that the reversal of linear compressibility at ~3 GPa was related to a structural transition involving the transfer of a proton from the framework's phosphonate group to the included water molecule:

$$R-PO_3H + H_2O \rightarrow R-PO_3^- + H_3O^+$$

Furthermore, in the material with a longer alkyl chain (ZAG-6), this was accompanied by a coiling of the organic linker, favorable at high pressure. This example of pressure-induced bond reorganization is something that is relatively rare in a soft porous crystal,[47,48] which can only be treated by computationally expensive quantum methods.

Another example is the study of the hydrothermal breakdown of flexible MOF MIL-53(Ga).[49] In the study of such a phenomenon, it is important to treat both the thermal effects and to allow a full liberty of the electronic degrees of freedom associated with bond breaking, and therefore the method of choice is *ab initio* molecular dynamics (AIMD; also called "first principles" molecular dynamics). Moreover, because the simulation times accessible to AIMD on periodic systems are relatively short (of the order of tens of picoseconds), and bond breaking is a rare event, we employed the metadynamics technique for free energy calculations along a carefully selected reaction coordination — in our case, related to the coordination of metal and linker. With this technique, we confirmed that the weak point of the MIL-53(Ga) structure is the bond between the metal center and the organic linker, and elucidated the mechanism by which the presence of water in the pores lowers the activation free energy for the breakdown. However, due to their high computational cost, full studies of flexible frameworks by *ab initio* MD are still relatively few and far between.[50,51,52]

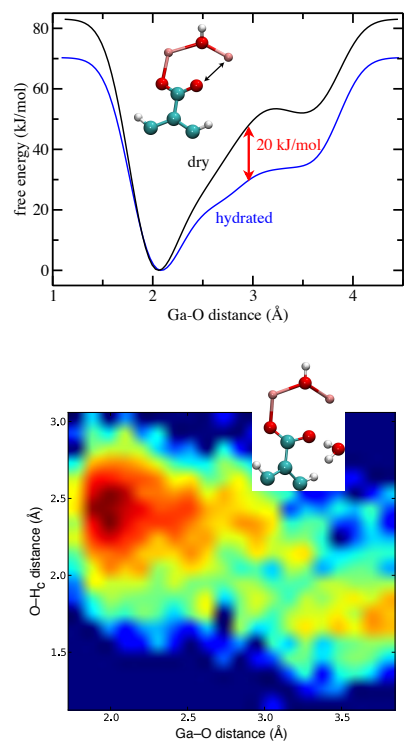

**Fig. 4** Top: Free energy profiles of Ga−O bond breaking for dry (black) and hydrated (blue) MIL-53(Ga) at 650 K. Bottom: 2D free energy profiles for hydrated MIL-53(Ga), as a function of Ga–O and O–H distances. Reproduced with permission from Ref. 49. Copyright 2015 American Chemical Society.

Finally, a third example of highly nonlinear phenomenon, this time upon adsorption of guest molecules, is the recent discovery of *negative gas adsorption*.[53] In this eye-catching case, the increase of gas pressure outside a flexible microporous MOF leads to a sudden contraction of the framework, accompanied by expulsion of a large fraction of the adsorbed gas — leading to a decrease in uptake upon increase of pressure, $(\partial N_{ads}/\partial P < 0)$, which is forbidden by thermodynamics. This transition was originally evidenced in material DUT-49,[54,55] a copper-based MOF built on long organic linkers containing a biphenyl unit at their center. By coupling in situ diffraction and spectroscopic methods as well as theoretical calculations at various scales, we have been able to provide microscopic insight into this transformation. We performed quantum chemistry calculations of the organic linker itself, under various constraints, as well as classical simulations of the thermodynamics of adsorption, and the dynamics of the framework.[56] We could thus show that the transition is associated with a buckling of the organic linker, triggered by the adsorption stress, leading to the shrinkage of the pores. Moreover, this buckling occurs only after a certain stress is reached, meaning that the system stays in a metastable state beyond the thermodynamic equilibrium, explaining the negative step in the adsorption isotherm. While this phenomenon is still relatively new, it appears already that other materials with the same topology can exhibit different behavior, depending on their linker length.[57]

## ■■ 3. Modelling of defects and disordered phases, and composite systems

Another important axis of our recent work involving soft porous crystals has been the drive to move beyond the view of these systems as defect-free, ordered single crystals of infinite size. The presence of defects and disorder, which are present to some extent in all solids, can be exacerbated in soft porous crystals and coupled to the presence of flexibility — the reason being a common root, in the entropy of these materials. This, in turn, is linked to the high dimensionality arising from the intramolecular degrees of freedom of the materials, the many corresponding 'soft' modes of low energy due to the relatively weak interactions involved in the assembly.[58]

### 3.1. Defects

Defects are present to some extent in all crystalline solids, and in several cases they are key to the properties or function of materials. Yet, the study of their occurrence in MOFs and their impact on their behavior is still relatively recent, and far from systematic. Experimental studies have largely focused on materials of the UiO-66 family, where defects are often present in relatively high numbers, and their concentration can be controlled by the synthesis conditions. The careful introduction of defects in MOFs (defect engineering, or so-called "defective by design" materials) is a very active topic of research, in particular for catalytic applications.[59,60,61,62]

On the topic of modelling of defects in soft porous crystals, two main axes of researched have been pursued. The first is the use of calculations, in synergy with experimental characterization techniques, to provide a better microscopic view of the MOF structure and the local defect sites. The exact atomistic details of defects, for example in UiO-66, has been largely debated in the past, and calculations of different structures have been proposed, depending on the nature of the terminating (or capping) groups.[63,64,65,66] density functional theory (DFT) calculations can be used to compare the formation energies of various configurations of a defect, either on cluster models or on fully periodic systems.

The second is the use of simulations in order to better understand the impact of the presence of defects on the properties of a given material. For example, the impact of missing linker defects on adsorption has been studied by several groups, showing the influence of their concentration and local distribution on both accessible surface area and pore volume. Such studies are typically performed in a classical approach, where the defect structure is assumed, based on experimental data or chemical intuition. Properties other than adsorption isotherms can also be calculated, however. We have shown, for example, that while $CO_2$ uptake is enhanced in zirconium-based UiO-66 by introduction of defects, this is accompanied by a reduction in mechanical stability due to the lower coordination of the framework (Figure 5).[67]

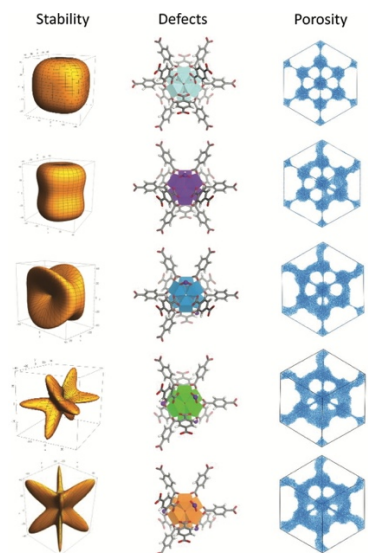

**Fig. 5** Direction Young's modulus (left) and available porosity (right) for UiO-66(Zr) materials with increasing number of missing linkers around a single metal node. Reproduced with permission from Ref. 67. Copyright 2016 Royal Society of Chemistry.

There are some cases, however, where the picture is not so grim. Lee et al. showed that it is possible to construct a multicomponent MOF with a "redundant" framework, where several linkers coexist in order to create a defect-tolerant material.[68] The quaternary MOF MUF-32 behaves in that way. It is built from a main load-bearing sublattice, in which additional linkers can be present but are not necessary for stability. In this way, high levels of vacancy defects can be introduced by their partial omission or removal, tuning the adsorption properties of the material without compromising its mechanical stability.

### 3.2. Disorder

The presence of defects, as described above, can induce disorder (partial or total) in MOFs — though it is far from being its only possible cause. Recent years have seen the emergence of studies on intrinsically disordered phases of MOFs, including the most amorphous phases (i.e., MOF glasses), as well as a few examples of studies on the distribution of defects.

Regarding the presence of disorder in crystalline MOFs, perhaps the best characterized system is that of UiO-66 materials with missing linker defects. It was extensively investigated, by both experimental and computational means, by the Goodwin and their collaborators. It was first shown that defects are not introduced in a random manner, but that correlated defect nanoregions emerge, whose size can be chemically controlled.[69] Further study showed that the inclusion of such a defective nanostructure could be used to tune the physical properties of thermally-densified UiO-66(Hf), creating colossal isotropic negative thermal expansion (NTE) — dependent on defect concentration.[70]

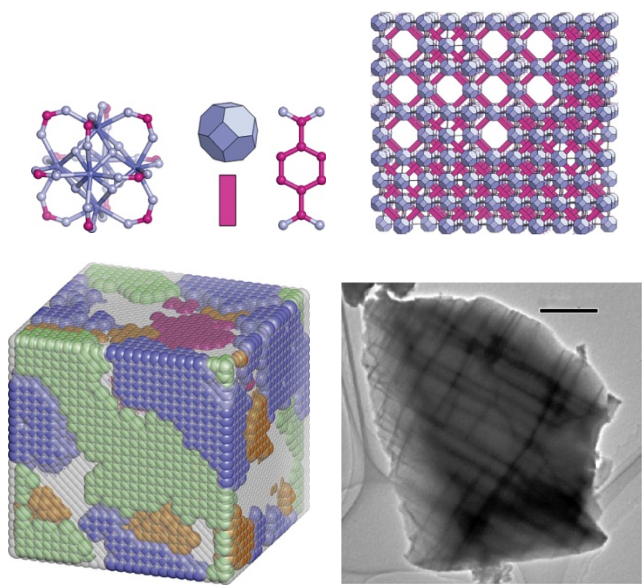

**Fig. 6** Top: Structural description of UiO-66(Hf) and a crystal with a defect-rich nanoregion. Bottom left: representation of defect-rich nanodomains, with four possible orientations in different colors. Bottom right: complex microstructure of UiO-66(Hf) crystallites observed experimentally. Reproduced from Ref. 69.

Another class of MOF systems of interest when it comes to the presence of disorder is that of heterometallic MOFs, *i.e.* framworks containing more than one type of metal center. In such systems, the distribution of metal centers is key to the properties of the system, with the two extreme cases being random mixing of metals or phase separation into pure-metal crystals or domains.[71] This is, from the point of view of molecular simulations, still an rather open question. Trousselet showed that the chemical nature of the metal cations, their relative sizes and the existence of charge transfer inside secondary building units are key in determining whether metal mixing is favorable in bimetallic UiO-66 and MOF-5.[72] Sapnik et al. used reverse Monte Carlo modelling to understand the distribution of metal centers in a mixed-metal Zn/Cd zeolitic imidazolate framework.[73] However, there is a scarcity of works dealing with the impact of this metal center disorder on MOF properties in general, and on flexible frameworks in particular.[74,75]

Finally, we note that while there have been many experimental studies of amorphous phases of MOFs[76,77,78,79] (for a brief review, see Ref. 80), they are for now unexplored from the computational point of view.

### 3.3. Composite systems

This section will be short, since there are relatively few papers that deal with this topic, but we felt the need to highlight the recent trend in modeling realistic systems at a higher scale (mesoscopic or macroscopic). In devices and applications, it is most likely that soft porous crystals will not be used in the form of isolated single crystals, but a nano- or micro-structured composite materials: thin films on a substrate, core–shell particles, mixed matrix membranes, etc. Therefore, efforts to model such complex systems, as well as MOF/polymer interfaces (for example) are necessary. The group of Maurin worked in that direction, describing at the microscopic level the interface between MOF crystal surface and polymer.[81] They studied the ZIF-8/PIM-1 interface (PIM = polymer of intrinsic microporosity), detailing the nature of the interface and the MOF/PIM interactions, and showing that the presence of the MOF surface impacts the polymer structure and dynamics at relatively long distances, up to 20 Å.[82] The limitations of a coarse-grained force field for the description of MOFs has meanwhile been tested by Dürholt et al. on the example of HKUST-1, with reasonable description of some (but not all) of the lattice dynamical features with only one coarse-grained bead for 30 atoms.[83] Finally, our own group has worked in a somewhat different direction, looking at two different descriptions of composite systems build from soft porous crystals and a polymer matrix. Using first a purely analytical mechanical description,[84] and then a macroscopic modelling approach by finite elements,[85] Evans studied the impact of composition and geometry on the macroscopic properties of nanostructures composites.

### ■■ 4. Perspectives

We have tried to give above a short account of the toolbox of theoretical approaches, developed in our group and others across the world, that have been used in order to study the behavior of soft porous crystals. This field is quite active and in constant development, so it is likely that by the time this account is published, novel simulation methodologies will have been published. Drawing perspectives is therefore akin to making predictions, which is difficult — especially about the future. However, it appears likely that some of the currently open questions will remain a challenge for a few more years at least. The description of defective and/or disordered systems is one of these challenges, where the length scales at play make computational approaches quite difficult. For a similar reason, the behavior of inhomogeneous systems, nanostructured composites, and polycrystalline systems is only rarely addressed in theoretical studies. Effects of crystal size will also need to be understood better, especially when we know that they can — at least in some cases — drastically affect the behavior of soft porous crystals[86,87,88] by enhancing or disabling flexibility.

Finally, while stimuli such as temperature, pressure and adsorption have been widely studied, the behavior of soft porous crystals under electric field, light, magnetic field, liquid-phase intrusion, pH or chemical gradients, … is still a largely open question. The influence of light on photoresponsive MOFs is of particular interest, and has been studied by several authors in the past few years.[89,90,91] However, the photophysical properties of MOFs have been less explored from a computational point of view,[92,93,94] especially so for soft materials with responsive behavior or large-scale flexibility. This can probably be attributed to the high computational cost of the quantum chemistry methods needed to study the excited states of these complex supramolecular solids.

## ■■ Acknowledgments

I thank the Japan Society of Coordination Chemistry for their International Award for Creative Work and the many fruitful discussions during the 68th Conference of JSCC in Sendai. I thank Alain Fuchs and Anne Boutin for our continuing collaboration on this exciting topic, including innumerable discussions and lots of fun projects, as well as all collaborators past and present for making work such a pleasure.

Finally, I thank the MOF2018 conference (held in Auckland, New Zealand) and Singapore Airlines for the long but pleasant flights during which this paper was written.

## ■■ References